\documentclass[12pt]{article}
\input{epsf}
\usepackage{amsfonts}

\usepackage{graphicx}
\usepackage{amsmath}

\input{tcilatex}
\def\spa#1{\phantom{\fbox{\rule[-#1cm]{0cm}{0cm}}}}
\def\mysquare{\nabla^2}
\begin{document}
\baselineskip16pt

\begin{titlepage}
\begin{flushright}
ITFA--2002--11\\
LPTENS--02--29\\
\end{flushright}
\vspace{8 mm}
\begin{center}
{\Large \bf A Resolution of the Cosmological}\\
\vspace{5 mm}
{\Large \bf Singularity with Orientifolds}\\
\end{center}

\vspace{8 mm}

\begin{center}

{\bf L. Cornalba}$\,^\dagger$,
{\bf M.S. Costa}$\,^\ddagger$\footnote{\,On
leave from Departamento de F\'\i sica,
Faculdade de Ci\^encias, Universidade do Porto.}
and
{\bf C. Kounnas}$\,^\ddagger$

\vspace{4mm}
{$\,^\dagger$ Instituut voor Theoretische Fysica, Universiteit van
Amsterdam}\\
{\small Valckenierstraat 65, 1018 XE Amsterdam, The Netherlands\\
lcornalb@science.uva.nl}\\
\vspace{5mm}
{$\,^\ddagger$ Laboratoire de Physique Th\'eorique, Ecole Normale
Sup\'erieure\footnote{Unit\'e mixte du CNRS et de l'Ecole Normale
Sup\'erieur, UMR 8549.}}\\
{\small 24 rue Lhomond, F-75231 Paris Cedex 05, France\\
miguel@lpt.ens.fr, kounnas@lpt.ens.fr}\\

\vspace{5mm}

\end{center}

\begin{abstract}
We propose a  new cosmological scenario which resolves the conventional
initial singularity problem.
The space--time geometry has an unconventional time--like singularity
on a lower dimensional hypersurface, with localized energy density.  
The natural interpretation of this singularity in string theory 
is that of negative tension branes, for example the orientifolds of
type II string theory. Space--time ends at the orientifolds, and it is
divided in three regions: a contracting region with a future
cosmological horizon; an intermediate region which ends at the
orientifols; and an expanding region separated from the
intermediate region by a past cosmological horizon.
We study the geometry near the singularity of the proposed
cosmological scenario in a specific string model. 
Using D--brane probes we confirm the interpretation of the brane
singularity as an orientifold. The boundary conditions on the
orientifolds and the past/future transition amplitudes are well
defined. Assuming the trivial vacuum in the past, we derive a thermal
spectrum in the future.
\end{abstract}

\end{titlepage}

\newpage

\section{Introduction}

Gravity is an attractive force. This is the basic reason for the
presence, in the standard cosmological models, of a space--like
big--bang singularity in the past. The existence of an initial
singularity is, to a large extent, independent of the model, and in
fact one can show that, under general assumptions, a singularity in
the cosmological past is inevitable \cite{HawkingEllis}. 
Despite these problems, the standard cosmological models (far) below
the Planck scale are extremely successful in explaining the present
experimental data.

Attempts to solve the cosmological singularity problem lead to the pre
big--bang scenario conjecture \cite{Veneziano:00}, motivated by the
existence of a minimum distance in string theory. However, either the exact
string backgrounds are not realistic \cite{KiriKoun}, or they show
time--like singularities  whose nature has remained unclear
\cite{Kounnas:92,Grojean:01,CC}. 
More recently, the  big--bang singularity was investigated in the framework 
of brane cosmology as collision of branes \cite{collision}. However, in
these scenarios the cosmological singularity is space--like and does not
have a clear interpretation as a brane.

The simplest way to avoid the cosmological singularity is to introduce
a bulk positive cosmological constant, which introduces a uniform
negative pressure. In the absence of other contributions to the
stress--energy tensor one obtains a de Sitter space--time.
A de Sitter cosmology is problematic both from the string theory point 
of view and from a phenomenological view--point. In string theory it
has not been possible, up to now, to construct a (meta)--stable string
background with a de Sitter geometry. From the cosmological point of
view it is difficult to connect a de Sitter phase to the usual matter
dominated universes which describe present day evolution of the
universe, and at the same time retain a singularity--free space--time. 

Although problematic, de Sitter space avoids the cosmological
singularity by introducing an effectively {\it repulsive} part to the
gravitational interaction due to the negative pressure induced by the
positive cosmological constant. Another natural way to avoid the
big--bang singularity is the introduction of a negative cosmological
constant localized on hyperplanes of lower dimension. These
lower dimensional branes have negative tension and therefore will fill
a repulsive gravitational force induced by standard matter in the
bulk. Even though it is
problematic in pure gravity to consider objects with negative mass,
they appear quite naturally in string theory. Orientifold planes are,
for example, hypersurfaces of arbitrary dimension, charged under the
Ramond--Ramond fields, with negative tension and no localized
dynamical degrees of freedom \cite{Polchinski:96,Angelantonj:02}. 
From the point of view of gravity,
O--planes act as sources of matter and gravity fields by adding a local
term in the supergravity action 
$$
|T|\int_\Gamma d^dx\,e^{-\phi}\sqrt{-\det G}\ \pm Q\int_\Gamma A_{d}\ ,
$$
on the $d$--dimensional hypersurface $\Gamma$, where $T$ and $Q$ are
the tension and the charge respectively. This acts as a negative
cosmological term localized on the surface $\Gamma$, together with the
coupling to the other supergravity fields. The geometry which
corresponds, in supergravity, to O--planes is similar to the one
describing D--branes, but has negative mass and develops a naked
time--like singularity. 

In this paper we will show that it is possible to resolve the
conventional cosmological singularity problem using orientifolds, i.e. by
introducing a negative cosmological constant localized on  
domain walls at the boundary of space--time. The basic idea is to
replace the big--bang singularity at $t=0$ by a cosmological horizon
and to continue space--time across this horizon. Examples of such
space--time causal structure were presented in \cite{Kounnas:92} for a
spatially flat cosmology that becomes isotropic at late times and in 
\cite{CC} for isotropic open cosmologies (with hyperbolic spatial
sections). Other examples were given in \cite{Grojean:01}. To understand
the existence of this horizon it is convenient to consider the
description of $(d+1)$--dimensional flat space--time using the Milne and
Rindler `polar' coordinates. Along the `cosmological' Milne wedge the
metric depends on the cosmological coordinate $t$, however, as one
crosses the horizon the metric dependents on a space--like coordinate
$x$, and it is foliated by $d$--dimensional de Sitter slices. The
idea proposed in \cite{CC}, is that one may start integrating the
classical equations of motion starting from this horizon regardless of the
theory under consideration, imposing the existence of a cosmological
horizon. In some cases this space--time causal structure arises as a
Ka\l u\.{z}a--Klein compactification of  flat space--time, which
can be described by a string theory orbifold construction, following
the earlier work of \cite{Khoury:01} (see also \cite{orbifold}).

In the specific model we shall consider, which can be embedded in Type
II string theory, the geometry in the Rindler patch develops a
time--like naked singularity, where the metric has the correct form to
be interpreted as an orientifold plane with a de Sitter world--volume
(see figure 1). This geometry describes an open universe with contracting
and expanding regions, together with an intermediate region
bounded by an orientifold plane. The latter region smoothly connects
the would be `big--crunch' and `big--bang' of conventional
cosmological models. This resolution of the cosmological singularity
allows the calculation of transition amplitudes from the contracting to
the expanding phases. In particular, we shall calculate the
past/future vacuum to vacuum amplitude for a free field propagating in
the cosmological background. Assuming a trivial vacuum at the
asymptotic past, we obtain a thermal spectrum in the far future. This
analysis is similar to that of Hawking in the derivation of  black
hole radiation \cite{Hawking}, but applied to our cosmology. 

\begin{figure}
\centering
\includegraphics{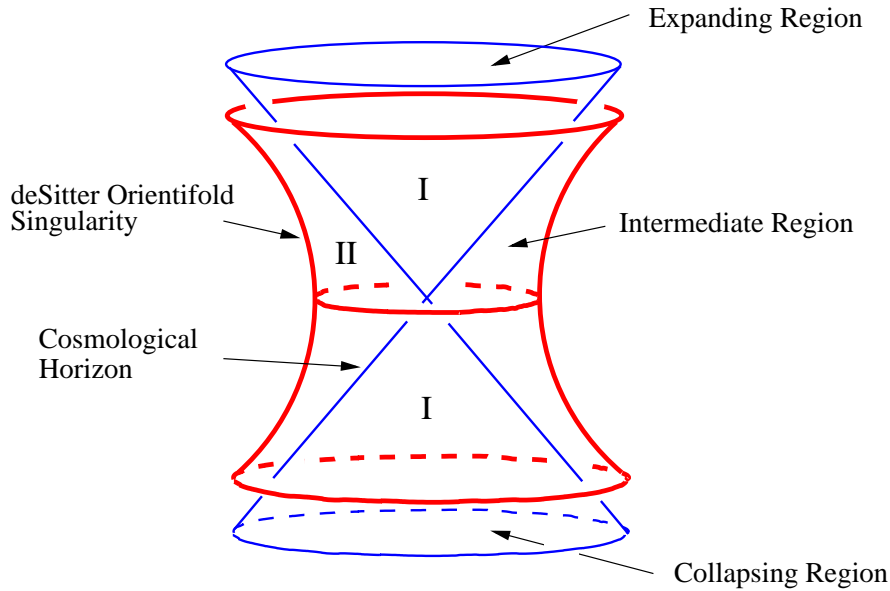}
\caption{\small{De Sitter orientifold as the boundary of space--time. We shall
see that the associated geometry develops a cosmological horizon that
allows the smooth transition form a contracting to an expanding cosmological
phase.}}
\end{figure} 

To describe dynamically the phases of the universe, we can think of a
large spherical charged brane as the boundary of the universe, which
is contracting due to the brane's tension. During the collapse, the
brane will back--react on the geometry and it will create, in the
center of the sphere, a positive energy density. If the brane has
positive tension, it will eventually collapse due to its own
gravitational and electric forces. On the other hand, if the brane has
negative tension, it will interact with its own gravitational and
electric fields with opposite sign, and this will invert the
contraction to an expanding phase. In this sense, localized negative
tension objects can be used to avoid the big--bang singularity. 

One may  ask at this point why the choice of orientifolds is justified
in cosmology. In our point of view this choice has a natural
explanation in  string theory. Indeed, to create a non--trivial
gravitation background with broken supersymmetry one has to start with
non--BPS configuration of branes. If not, the vacuum solution  will be
stable and stationary due to the supersymmetry. One possibility is to
construct type II models where the RR charge is canceled only with
orientifold planes $O-\overline{O}$ localized in two different points, and
without any $D$ or $\overline{D}$ branes. Models of this kind are
constructed in the literature \cite{Orientifold,Angelantonj:02}. This
configuration breaks
supersymmetry, and it is incompatible with a flat space--time and
constant dilaton. The induced geometry in the presence of $O-\overline{O}$
orientifolds is precisely our cosmological solution. As we shall show
in section 3, the $O-\overline{O}$ configuration is stable. There is an
attractive electric force which is  balanced by an  effective
repulsive force created by the gravitational back--reaction of the
orientifolds. Thus, localized negative tension objects can be used to
avoid the conventional big--bang singularity. 

Another starting point, is to dress the orientifolds with branes
and to consider the configuration $OD-\overline{OD}$.
This configuration is consistent classically  with a flat space--time
and constant dilaton. However, at the quantum level it is unstable due
to the lack of supersymmetry. An  effective potential is created
bringing the branes and anti--branes together which annihilate each
other. After this annihilation, we are left with the  
$O-\overline{O}$ orientifolds, the background fields are no longer
flat and give rise to the cosmological scenario described above. We
shall confirm the structure of the effective potential among branes
and orientifolds by probing the geometry with D--branes. Since
orientifolds are a source of the gravitational and RR fields but do
not have local degrees of freedom, the acceleration of the universe
can be seen effectively from the condensation of the tachyon field
associated to the above process. Indeed, tachyon driven
time--dependent string backgrounds have been the subject of recent
investigation \cite{tachyon} (see \cite{timedep} for related work). 

\section{A Simple Cosmological Solution}

We start quite generally by considering $\left( d+1\right) $--dimensional
gravity coupled to a scalar field $\psi $
\begin{equation}
S=\frac{1}{2\kappa^2{\mathcal E}^{d-1}}\int d^{d+1}x\sqrt{|g|}
\left[R-\frac{\beta}{2}\left(\nabla\psi\right)^{2}
-V\left(\psi\right)\right]\ , 
\label{eq100}
\end{equation}
where ${\mathcal E}$ is an energy scale. 
In this equation, $V\left(\psi\right)$ is the potential for the scalar
field, $\beta$ is a dimensionless constant and the coordinates are
dimensionless in units of $1/{\mathcal E}$. Following \cite{CC}, we
are interested in cosmological solutions to the equations of motion of
(\ref{eq100}) with a contracting and an expanding phase (which we call
region $I$), together with an intermediate region (denoted by region
$II$). The expanding phase is the standard
Robertson--Walker geometry for an open universe 
\begin{equation}
\begin{array}{rcl}
\displaystyle{ds_{d+1}^{\ 2}} 
&=&
\displaystyle{-dt^{2}+a_{I}^{\ 2}
\left(t\right)ds^{2}\left(H_{d}\right)}\ ,
\spa{0.2}\\
\psi &=&
\displaystyle{\psi _{I}\left(t\right)}\ ,  
\end{array}
\label{eq10}
\end{equation}
where $H_{d}$ is the $d$--dimensional hyperbolic space with unit radius.
Therefore, the dynamics of the system is described by the scalar and
Friedman equations 
\begin{equation}
\begin{array}{rcl}
\displaystyle{\ddot{\psi}_{I}+d\,\dot{\psi}_{I}\,\frac{\dot{a}_{I}}{a_{I}}} 
&=&
\displaystyle{-\frac{1}{\beta}\,\frac{\partial V}{\partial \psi_I}}\ ,
\spa{0.5}\\
\displaystyle{\left(\frac{\dot{a}_{I}}{a_{I}}\right)^{2}
-\frac{1}{a_{I}^{\ 2}}} 
&=&
\displaystyle{\frac{1}{d\left(d-1\right)}
\left[\frac{\beta}{2}\,\dot{\psi}_{I}^{2}
+V\left(\psi_{I}\right)\right]}\ , 
\end{array}
\label{eq20} 
\end{equation}
where dots denote derivatives with respect to the cosmological time $t$.
The contracting phase is nothing but the time--reversed solution,
where we replace in (\ref{eq10}) $t$ by $-t$.

In order to have an intermediate region, we are interested in
solutions of (\ref{eq10}) where $a_{I}(t)$ and $\psi_{I}(t)$ are,
respectively, odd and even functions of $t$ with initial conditions 
\begin{equation}
a_{I}\left(t\right)=t+\mathcal{O}\left(t^{3}\right)\ ,
\ \ \ \ \ \ \ \ \ \ \ \ \ \ \ \ \ 
\psi _{I}\left(t\right)=\psi_{0}+\mathcal{O}\left(t^{2}\right)\ ,  
\label{eq50}
\end{equation}
for small cosmological time $t$. This implies that the $t=0$ surface does
not correspond to a big--bang singularity, but represents a null
cosmological horizon, as described in \cite{CC}. In this case, the
space--time can be extended across the horizon to an intermediate region 
$II$, where the solution takes the form 
\begin{equation}
\begin{array}{rcl}
\displaystyle{ds_{d+1}^{\ 2}} 
&=&
\displaystyle{a_{II}^{\ 2}\left(x\right)ds^{2}\left(dS_{d}\right)+dx^{2}}\ , 
\spa{0.2}\\
\psi &=&\displaystyle{\psi_{II}\left(x\right)}\ ,
\end{array}
\end{equation}
with $dS_{d}$ the $d$--dimensional de Sitter space. The equations of
motion which determine $a_{II}$ and $\psi _{II}$ are again the
equations (\ref{eq20}), with the potential $V$ replaced by $-V$. 
To extend the solution from region $I$ to region $II$ it is convenient
to write the hyperboloid and de Sitter metrics in conformal coordinates
$(i=1,\cdots,d-1)$
\begin{equation*}
ds^{2}\left(H_{d}\right)=\frac{dy^idy_i+dy_d^{\,2}}{y_d^{\,2}}\ ,
\ \ \ \ \ \ \ \ \ 
ds^{2}\left(dS_{d}\right)=\frac{-dy_0^{\,2}+dy^idy_i}{y_0^{\,2}}\ .
\end{equation*}
Then, if we consider $a_{I}$ and $\psi _{I}$ as complex functions, the
solution can be extended by analytic continuing $t\to ix$, $y_d\to iy_0$,
together with   
\begin{equation*}
a_{II}\left(x\right)=-ia_{I}\left(ix\right)\ ,
\ \ \ \ \ \ \ \ \ \ \ \ \ \ \ \ \ \ \ \ 
\psi_{II}\left(x\right)=\psi_{I}\left(ix\right)\ .
\end{equation*}
Equivalently, given the boundary conditions (\ref{eq50}), regions $I$
and $II$ can be connected along the null cosmological horizon, just as
a Milne universe can be glued to a Rindler edge to form flat Minkowski
space. Moreover, it is clear that the solution possesses a global 
$SO\left(1,d\right)$ symmetry, which acts both on $H_{d}$ and on
$dS_{d}$. 

The existence of an intermediate region demands generically a field
theory effective potential whose form is highly restrictive due to the
unnatural (fine--tuned) boundary conditions. In the next section we
see that string effective supergravity theories provide a
Liouville--Toda like potential \cite{Antoniadis:86} which naturally
respects the boundary conditions at the horizon. 

\section{Embedding in String Theory}

Now let us consider a particular case of the construction of the previous
section which can be embedded in Type II string theory. For simplicity
we consider backgrounds with a non--trivial RR field. The
corresponding ten--dimensional Type II effective action has the form
($l_s=2\pi\sqrt{\alpha'}$)
\begin{equation*}
S=\frac{2\pi}{l_s^{\ 8}}
\left[\int d^{10}x\sqrt{|g|}e^{-2\phi}
\left(R+4\left(\nabla\phi\right)^{2}\right)
-\frac{1}{2}\int F\wedge\widetilde{F}\right]\ ,
\end{equation*}
where $F$ is the RR $\left( d+1\right) $--form field strength and 
$\widetilde{F}=\star F$ is the dual $\widetilde{d}$--form field
strength, with $\widetilde{d}=9-d$. A family of solutions,
parameterized by an arbitrary constant $g_s$, can be constructed by
considering the following ansatz\footnote{The
constant $g_s$ is related to the constant decoupled scalar field
$\rho$ in \cite{CC} by 
\begin{equation*}
g_s =\exp \left[\frac{d}{8}\frac{9-d}{3d-2}\,\rho\right]\ .
\end{equation*}Moreover, $\mathcal E$ is related to $E$ of 
\cite{CC} by $E^d = g_s {\mathcal E}^d$.} 
\begin{equation}
\begin{array}{c}
\displaystyle{{\mathcal E}^2\,ds^{2}=
\Lambda^{-\frac{1}{2}\frac{d+1}{d-1}}\,ds_{d+1}^{\ 2}
+\Lambda^{\frac{1}{2}}\,ds^{2}(\mathbb{E}^{\widetilde{d}})}\ ,
\spa{0.4}\\
\displaystyle{e^{\phi}=g_s\,\Lambda^{\frac{4-d}{4}}\ ,
\ \ \ \ \ \ \ \ \ 
\widetilde{F}=\frac{1}{g_s{\mathcal E}^{\widetilde{d}-1}}
\,\epsilon(\mathbb{E}^{\widetilde{d}})}\ ,
\end{array}
\label{eq1000}
\end{equation}
where we have defined 
\begin{equation*}
\Lambda=e^{2\beta\,\psi}\ ,
\end{equation*}
and the line element $ds_{d+1}$ is defined in the previous section. 
The physical interpretation of the constants $\mathcal E$ and $g_s$
will become clear below. Following \cite{CC}, the equations of motion
for the scale factor and scalar field are those presented
in the previous section with 
\begin{equation*}
V=\frac{1}{2}\,e^{-\psi }\ ,
\end{equation*}
and 
\begin{equation*}
\beta =\frac{1}{2}\,\frac{d-1}{3d-2}\ .
\end{equation*}

In \cite{CC} it was shown that, in the intermediate region $II$, the
geometry develops a time--like curvature singularity. The
behavior of the scalar field and scale factor near the singularity 
was seen to be
\begin{eqnarray}
a_{II}\left(x\right) &=&
a_s\left(x_s-x\right)^{\gamma}\ ,
\\
\psi_{II}\left(x\right) &=&
\log{\left[\eta\left(x_s-x\right)\right]^2}\ ,
\label{eq30}
\end{eqnarray}
where the constants $\gamma$ and $\eta$ read
\begin{equation*}
\gamma=\frac{2\beta}{d-1}\ ,
\ \ \ \ \ \ \ \ \ \ \ \ \ \ \ \ \ \ \ \ 
\eta^{-2}=4\beta\left(1-d\,\gamma\right)\ .
\end{equation*}
The dimensionless constant $a_s$ depends on the boundary conditions
imposed at the horizon; we shall come back to this issue later. 
The value of $x_s$, where $a\rightarrow 0$ and $\psi\rightarrow-\infty$,
can be written as
\begin{equation*}
x_{s}=e^{\frac{1}{2}\psi_{0}}c_{s}\ ,
\end{equation*}
where $\psi_{0}$ is the value of the scalar field at the horizon.
The constant $c_{s}$ depends only on the dimension $d$ of space and
is given numerically by 
\begin{table}[h]
\begin{center}
\begin{tabular}{ c|c c c c c c c} 
$\spa{0.1}d^{\phantom{\frac{2}{2}}}$ & $2$ & $3$ & $4$ & $5$ & $6$ & $7$ & $8$
\\ \hline
$\spa{0.1}c_s^{\phantom{\frac{2}{2}}}$ & $1.6$ & $1.9$ & $2.1$ & $2.3$ & $2.5$ & $2.6$ 
& $2.8$
\end{tabular}
\end{center}
\end{table}

\vspace{-0.7cm}
From the behavior of the scalar field at the singularity
$(\psi\rightarrow-\infty)$, at the horizon $(\psi=\psi_0)$ and at the
asymptotic future $(\psi\rightarrow+\infty)$, and from the fact that the
scalar field is a growing function as one moves from the singularity
to the horizon and from there to the asymptotic future, 
it is natural to use $\Lambda$ as a radial variable instead of $x$. 
At the singularity we have $\Lambda=0$  and at the lightcone 
$\Lambda=\Lambda_0\equiv e^{2\beta\psi_0}$. Then near the singularity,
in the limit $\Lambda\ll\Lambda_0$, we obtain the following
ten--dimensional background fields 
\begin{equation*}
\begin{array}{c}
\displaystyle{\mathcal{E}^2ds^2 = 
\Lambda^{-\frac{1}{2}}\,\mu\,ds^2(dS_d)
+\Lambda^{\frac{1}{2}}
\left[d\Lambda^2+ds^2(\mathbb{E}^{\widetilde{d}})\right]}\ ,
\spa{0.5}\\
\displaystyle{e^{\phi}=g_s\Lambda^{\frac{4-d}{4}}\ ,
\ \ \ \ \ \ \ \ \ \ \ \ \ 
F = -\frac{1}{g_s\mathcal{E}^d}\,d\Lambda^{-1}
\wedge\left[\sqrt{\mu^d}\,\epsilon(dS_d)\right]}\ ,
\end{array}
\end{equation*}
where $\mu$ is a constant determined from the constant $a_s$ in the
expansion (\ref{eq30}). If we substitute the de Sitter space $dS_d$ by
the Minkowski space $\mathbb{M}^d$ we obtain naively the
solution for D$p$--branes $(p=d-1)$ localized at $\Lambda=0$ and uniformly
smeared along $\mathbb{E}^{\widetilde{d}}$. The usual harmonic
function $H$ is simply $H=\Lambda$ because there a unique transverse
direction. However, the tension associated with the harmonic function
proportional to $-\bigtriangleup H$ is negative. Therefore the
singularity is correctly interpreted as non--dynamical
orientifold $(d-1)$--planes smeared along the
$\mathbb{E}^{\widetilde{d}}$. When $\Lambda\rightarrow 0$
the radius of the de Sitter slice $\sqrt{\mu}\,\Lambda^{-1/4}$
diverges, the orientifold looks flat and one approaches the usual
supersymmetric solution. This solution is interpreted as the geometry
of a de Sitter O$p$--plane ($p=d-1$) with radius $R_{dS}\sim 1/\mathcal{E}$
and delocalized along $\mathbb{E}^{\widetilde{d}}$. Below we shall
compute the number of orientifold planes in terms of the parameters of
the solution. 

To understand the behavior of the geometry away from the orientifold
planes it is useful to keep using the function $\Lambda$ as the radial
coordinate. Then the solution (\ref{eq1000}) becomes
\begin{equation}
\begin{array}{c}
\displaystyle{\mathcal{E}^2 ds^{2}
=\frac{G}{\Lambda^{\frac{1}{2}}}\,\mu\,ds^{2}\left(dS_{d}\right)
+\frac{\Lambda^{\frac{1}{2}}}{H}d\Lambda^{2}
+\Lambda^{\frac{1}{2}}ds^{2}(\mathbb{E}^{\widetilde{d}})}\ , 
\spa{0.6}\\
\displaystyle{e^{\phi }=g_s\Lambda ^{\frac{4-d}{4}}\ ,
\ \ \ \ \ \ \ \ \ \ \ \ \ \ \ 
F=\frac{1}{g_s\mathcal{E}^{d}}\,\frac{1}{\Lambda^{2}}\sqrt{\frac{G^{d}}{H}}
\,d\Lambda\wedge\left[\sqrt{\mu^d}\,\epsilon(dS_d)\right]}\ ,
\end{array}
\label{eq2000}
\end{equation}
where $G(\Lambda)$ and $H(\Lambda)$ are dimensionless functions defined by 
\begin{equation*}
\mu\, G=\Lambda^{-\frac{1}{d-1}}\,a_{II}^{\ 2}(\Lambda)\ ,
\ \ \ \ \ \ \ \ \ \ \ \ \ \ \ \ \ \ \ 
H=\Lambda^{\frac{d}{d-1}}\left(\frac{d\Lambda}{dx}\right)^{2}\ .
\end{equation*}
Notice that, to match the near singularity behavior,  
the function G satisfies $G(0)=1$. 
Naively the solution is parameterized by $g_s$, $\mathcal{E}$ and the
position of the horizon $\Lambda_0$. However, we can set
$$
\Lambda_0= 1\ ,
$$ 
using the rescaling of the coordinates 
$\Lambda\to c\Lambda$  and 
$\mathbb{E}^{\widetilde{d}}\to c\,\mathbb{E}^{\widetilde{d}}$,
which leaves the form of the solution invariant if we also redefine
$g_s\to  c^{(d-4)/4}\,g_s\,$,  
$\mathcal{E}\to c^{5/4}\,\mathcal{E}$ and 
$G(\Lambda)\to c^3 G(\Lambda)$, $H(\Lambda)\to H(\Lambda)$.

With this rescaling the parameters $g_s$ and $\mathcal{E}$ are 
the string coupling $e^{\phi }$ and the electric field
$g_s\sqrt{-F^2}$ at the core of the solution, i.e. at the cosmological
horizon. In terms of these parameters we can compute the number of
orientifold planes. To this end we compactify the transverse
space on a torus $\mathbb{T}^{\widetilde{d}}$ with proper volume at
the horizon 
\begin{equation*}
V = \mathcal{E}^{-\widetilde{d}}
\int_{\mathbb{T}^{\widetilde{d}}}
\epsilon(\mathbb{E}^{\widetilde{d}})\ .  
\end{equation*}
Then the number of O--planes is given by
\begin{equation*}
N=\frac{2^{\,6-d}}{l_s^{\ 8-d}}
\int_{\mathbb{T}^{\widetilde{d}}}\widetilde{F}
=2^{\,6-d}\,\,
\frac{l_s\mathcal{E}}{g_s}\,\frac{V}{l_s^{\ \widetilde{d}}}\ .
\end{equation*}

Now let us analyze the behavior of the solution near $\Lambda =1$
and near $\Lambda=0$. It is easy to show, from the boundary conditions 
(\ref{eq50}) on the cosmological horizon, that 
\begin{equation}
G\left(1\right)=H\left(1\right)=0\ .  
\label{eq4000}
\end{equation}
To analyze analytically the behavior close to the singularity, we
need to rewrite the scalar and Friedman equations in terms of $G$ and
$H$ as functions of $\Lambda$. It is tedious but not difficult to show
that equations (\ref{eq20}) are equivalent to 
\begin{equation}
\begin{array}{rcl}
\displaystyle{\left(\frac{G'}{G}\right)^2+\frac{2}{(d-1)\Lambda}\,\frac{G'}{G}}
&=& 
\displaystyle{\frac{4\Lambda}{\mu GH}
+\frac{2}{d(d-1)\Lambda^{2}}\left(1-\frac{1}{H}\right)}\ ,  
\spa{0.7}\\
\displaystyle{\frac{H'}{H}+d\,\frac{G'}{G}}
&=&
\displaystyle{\frac{2}{\Lambda}\,\left(1-\frac{1}{H}\right)}\ ,
\end{array}  
\label{eq3000}
\end{equation}
where $'$ denotes derivative with respect to $\Lambda$.
Then we see that, to keep the behavior of the solution near the
orientifold at $\Lambda =0$, we must have 
\begin{equation*}
H\left( 0\right) =1\ .
\end{equation*}
On the other hand, when solving the equations starting from the
singularity, we are free to choose the initial condition
\begin{equation*}
G'\left( 0\right) =-\sigma\ ,
\end{equation*}
for arbitrary $\sigma$. 
Then, equations (\ref{eq3000})
can be solved in power series in $\Lambda $ to give 
\begin{equation}
\begin{array}{l}
\displaystyle{G\,=\,1-\frac{\sigma}{\mu}\,\Lambda-
\frac{\sigma^2}{\mu^2}\,\frac{d-1}{2}\,\Lambda^{2}+\cdots}\ ,
\spa{0.5}\\
\displaystyle{H\,=\,1-\frac{\sigma}{\mu}\,d\,\Lambda 
+\frac{\sigma^2}{\mu^2}\,\frac{d\left(d-1\right)}{2}\Lambda^{2}+\cdots}\ .  
\end{array}
\label{eq6000}
\end{equation}
Notice that for $d=1$ the solution can be found exactly \cite{CC}
because the geometry arises as a quotient of flat space and the
functions $G$ and $H$ are linear in $\Lambda$. 

We have seen that, starting from the orientifold singularity at 
$\Lambda=0$ we have two degrees of freedom $\mu$ and
$\sigma$ in the solution of (\ref{eq3000}). On the other hand,
starting from the light--cone we do not have any freedom. Therefore
the solutions that develop a cosmological horizon (\ref{eq4000})
correspond to a special value of $\mu$ and $\sigma$. For the
cosmological solutions we are analyzing, these constants 
depend only on the spatial dimension $d$ and are given numerically by  
\newpage
\begin{table}[h]
\begin{center}
\begin{tabular}{ c|c c c c c c c c } 
$\spa{0.2}d^{\phantom{\frac{2}{2}}}$ & $1$ & $2$ & $3$ & $4$ & $5$ & $6$ & $7$ & $8$
\\ \hline
$\spa{0.2}\mu^{\phantom{\frac{2}{2}}}$ & $1$ & $4.5$ & $5.2$ & $5.9$ & $6.7$
& $7.4$ & $8.2$ & $8.9$
\\ 
$\spa{0.2}\sigma^{\phantom{\frac{2}{2}}}$ & $1$ & $3.1$ & $2.9$ & $2.9$ & $2.9$ 
& $2.9$ & $2.9$ &$2.9$
\end{tabular}
\end{center}
\end{table}
\vspace{-0.6cm}
For large cosmological times in region $I$ the geometry is that of a
curvature dominated universe.

\subsection{Probing the Geometry with Branes}
 
Now that we have seen the string theoretic origin of the cosmological
singularity and how the geometry behaves as we move away from the
orientifolds, we would like to study the motion of D--branes in this
geometry. We are going to consider a spherical brane on the de Sitter
slices $dS_{d}$, positioned in region $II$ at some fixed value of
$\Lambda<\Lambda_{0}$, and we are going to compute the static potential
$\mathcal{V}\left(\Lambda\right)$ seen by the brane. Notice that by
static potential we mean the potential as seen by an observer on the
brane, i.e. an observer placed at constant $\Lambda$. 

It is well known that, in the case of flat branes, there is a
brane/anti--orientifold repulsion, whereas there is no force between
parallel branes and orientifolds (of opposite charge), due to the BPS
nature of the configuration. 
In the present case, we will see that the presence of a
de Sitter world volume of both the source and the probe brane induces
a repulsive force also on the probe branes with charge opposite to
that of the orientifold singularity. The force, when the
probe is placed close to the source, is \textit{repulsive} and
tends to move the probe brane away from the singularity. 
Bellow we shall argue that the force inducing the brane probe
to collapse is partially gravitational in character and arises from
the energy density of the fields generated by the de Sitter
orientifolds. This same energy density will act as an
anti--gravitational force on the orientifolds preventing them from
collapse. This mechanism provides a smooth transition from a
big--crunch to a big--bang phase, which is regular in string theory
and should allow the computation of transition amplitudes from the
contracting to the expanding region. Section four gives the simplest
example of such computation. 

The static potential for a spherical D$p$--brane ($p=d-1$) in the cosmological
background can be deduced starting from the metric (\ref{eq2000}). An
easy calculation shows that the Born--Infeld and the Wess--Zumino
pieces  give a total action 
\begin{equation*}
\mathcal{V}_{\pm}\left(\Lambda\right) 
=\frac{1}{\Lambda}\sqrt{G^d}
\mp\int\frac{d\Lambda}{\Lambda^{2}}\sqrt{\frac{G^d}{H}}\ .
\end{equation*}
The potential $\mathcal{V}_{+}$ ($\mathcal{V}_{-}$) is the potential seen by
a probe with charge equal (opposite) to those of the orientifolds forming the
singularity. Using the solution (\ref{eq6000}) we easily see that 
\begin{equation*}
\mathcal{V}_{+}\left(\Lambda\right)=\frac{2}{\Lambda }
+\cdots\ ,
\end{equation*}
and therefore $\mathcal{V}_{+}$ represents, as expected, a strongly
repulsive potential. As we discussed above, the potential $\mathcal{V}_{-}$
is more interesting physically, and vanishes for the case of parallel flat
branes and orientifolds. Again using (\ref{eq6000}) one can show that 
\begin{equation*}
\mathcal{V}_{-}\left(\Lambda\right)
=-\frac{d}{2}\,\frac{\sigma }{\mu}
-\frac{d\left(3d-2\right)}{8}\,\frac{\sigma^{2}}{\mu^{2}}\,\Lambda
+\cdots\ .
\end{equation*}
Note that, as long as $\sigma \neq 0$ (as it is the case for the
cosmological solutions), the potential decreases as a function of
$\Lambda$ for any value of $d\geq 1$, and therefore there is a
repulsive force which moves the probe away from the singularity at
$\Lambda=0$. One could think that this repulsive force is only due to
the tension of the spherical brane probe. On the other hand, the
analysis of the $d=1$ case, where the tension is absent, shows that
this intuition is not entirely correct. 

Let us consider a $O-\overline{O}$ pair in flat space at some large
distance and a $D$ particle near the $O$ plane. The only force
on the $D$ particle is the gravitational and electric repulsive
force due to the $\overline{O}$ plane, since a $OD$ pair is a BPS
configuration. Now let us turn on gravity and consider the
back--reaction on the geometry due to the presence of the
orientifolds. For this case we can define a coordinate $y$ by 
$\Lambda=1-y^4$, such that $y\in ]-1,0[$ corresponds to the part of
region $II$ with the $O$ plane, and $y\in ]0,1[$ corresponds to the
part with the $\overline{O}$ plane. We recall that $y=0$ is the
horizon where this coordinate system becomes singular. Then, recalling
that $G=H=1-\Lambda$, the potential
in the whole of region $II$ is given by (see figure 2)
$$
\mathcal{V}_D(y)=\frac{y^2}{1-y^2\,{\rm sign}(y)}\ .
$$
The fields generated by the orientifolds  induce an energy
density between them, which interacts gravitationally with
the $D$--particle. Indeed, this computation shows that the latter
gravitational attraction wins over the $D-\overline{O}$
repulsion. We conclude that the force that brings the $D$ particle
away from the $O$--plane is due to the back--reaction on the geometry,
which is creating an extra energy density that interacts
gravitationally with the probe. 

\begin{figure}
\centering
\includegraphics[width=6cm]{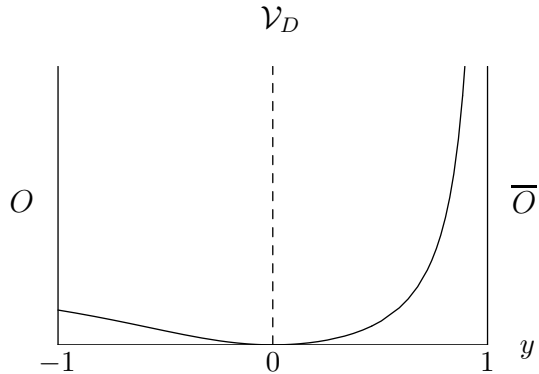}
\put(-90,120){$\mathcal{V}_D$}
\put(-185,50){$O$}
\put(5,50){$\overline{O}$}
\put(-174,-10){\small{$-1$}}
\put(-88,-10){\small{$0$}}
\put(-7,-10){\small{$1$}}
\put(8,-3){\small{$y$}}
\caption{\small{The potential $\mathcal{V}_D$ seen by a $D$--brane probe in the
geometry created by the $O-\overline{O}$ system, for $d=1$. The potential
repels the $D$--brane from both the $O$ and the $\overline{O}$--plane. The
vertical dashed line represents the location of the cosmological horizon.}}
\end{figure}

This energy density induced by the orientifolds is also responsible
for the repulsion between the 
O--planes, explaining the structure of the geometry against the naive
expectation that the $O-\overline{O}$ system is unstable under collapse and
annihilation. To confirm this interpretation let us show that the
electric field at the center of the exact solution is larger than the
electric field generated by the $O-\overline{O}$ system if we had neglected
back--reaction. This shows that the extra energy density is concentrated
at the center of the geometry. At the center, which is at a proper
distance $L$ from the orientifolds  given by
$$
\mathcal{E}L\,=
\int_0^1\frac{\Lambda^\frac{1}{4}}{\sqrt{1-\Lambda}}d\Lambda\,
=\frac{2\sqrt{2}}{21\sqrt{\pi}}
\,\Gamma\left[\,1/4\,\right]^2\,
\simeq\,1.75\ ,
$$
the electric field of the exact solution couples to gravity with strength
$\mathcal{E}$.  On the other hand, in flat space--time and
assuming linear superposition, the electric field of a $O-\overline{O}$
system at distance $2L$ is given in the intermediate point by  
$$
\frac{8}{5\,L}<\mathcal{E}\ ,
$$
thus confirming our claim.

In the general case, the $O-\overline{O}$ pair is really a spherical
orientifold and the energy density around the center of the geometry
gravitationally attracts the brane probes. Moreover, the same 
energy density prevents the orientifold sphere from collapsing, and is
actually the cause of the transition from the contracting to the
expanding phases.  

\section{Thermal Radiation from the Past Cosmological Vacuum}

We wish to analyze the propagation of matter in the cosmological
backgrounds studied in the previous sections. For simplicity, we shall
consider the case $d=1$ which arises \cite{CC} as an orbifold 
of Minkowski space under a combined boost and translation.

Let us consider a scalar field $\Xi$ of mass $m$ in
three--dimensional Minkowski space ${\mathbb M}^3$  with coordinates 
$X^\pm=T\pm X$ and $Y$, which satisfies the usual
Klein--Gordon equation 
\begin{equation*}
\left(-\mysquare + m^2\right) \Xi = \left(4 \partial_+\partial_- -
\partial^{\ 2}_Y + m^2\right) \Xi = 0\ .
\end{equation*}
We want to consider solutions of this equation which are invariant
under the identification $\Gamma$ defined by 
$X^\pm \sim e^{\pm 2\pi\Delta} X^\pm$ and
$Y\sim Y+ 2\pi R$. We conveniently define the energy scale 
$E$ to be $E=\Delta/R$. These identifications are
trivially satisfied with the separation of variables
\begin{equation*}
\Xi\left(X^+,X^-,Y\right)=\chi\left(X^+,X^-\right)e^{-ipEY}\ ,
\end{equation*}
where $\chi$ satisfies
\begin{equation}
\chi\left(e^\sigma X^+,e^{-\sigma} X^-\right) = e^{iJ\sigma}
\chi\left(X^+,X^-\right)\ ,
\label{spin}
\end{equation}
for any $\sigma$ and where
\begin{equation*}
J=p+\frac{n}{\Delta}\ .
\end{equation*}
The integer $n$ is the Ka\l u\.{z}a--Klein charge, and from now on we shall
consider the uncharged $n=0$ case.
In this case the Klein--Gordon equation becomes
\begin{equation}
\left( 4 \partial_+\partial_- + (\omega E)^2 \right) \chi = 0\ ,
\label{KG}
\end{equation}
where
$$
\omega^2 = \left(\frac{m}{E}\right)^2 + p^2\ .
$$
It is clear that $\chi$ satisfies the Klein--Gordon equation in
two--dimensional flat space for a particle of effective mass 
$\omega E$
and lorentzian spin $J$. Such equation is naturally solved by
introducing lorentzian polar coordinates in the Milne and Rindler
edges of the two--dimensional flat space--time, which correspond
respectively to regions $I$ and $II$ of the cosmological solution. The
polar coordinates read
\begin{equation}
\begin{array}{ll}
{\rm Region\ }I:\ \ \ \ \ \ \ \ &
EX^\pm = t\, e^{\pm z}
\spa{0.1}\\
{\rm Region\ }II:&
EX^\pm = \pm x\, e^{\pm s}
\end{array}
\label{polar}
\end{equation}
and the boost (\ref{spin}) acts as spatial translation along $z$ and time
translation along $s$ in the two regions, respectively. Therefore the
linearly independent solutions to (\ref{KG}) are (see \cite{Tolley:02}
and  references therein)
\begin{equation*}
\begin{array}{ll}
{\rm Region\ }I:\ \ \ \ \ \ \ \ &
e^{ipz}\,\mathcal{H}_p(\omega t)\ ,\ \ \ \ \ \ \ \ \ \,e^{ipz}\,\mathcal{H}_p(-\omega t)
\spa{0.2}\\
{\rm Region\ }II:&
e^{ips}\,\mathcal{H}_p(i\omega |x|)\ ,\ \ \ \ \ \ e^{ips}\,\mathcal{H}_p(-i\omega |x|)
\end{array}
\end{equation*}
where $\mathcal{H}_p$ is a properly normalized Hankel function 
$$
\mathcal{H}_p(v) = \sqrt\frac{\pi}{2} e^{-\frac{\pi}{2}|p|}\, 
H^{(1)}_{i|p|}(v)\ ,
$$
with the property that 
$$
\overline{\mathcal{H}_p(v)} = \pm \mathcal{H}_p(-\overline{v})\ ,
\ \ \ \ \ \ \ \ \ 
\pm{\rm Im}(v)<0\ ,
$$
and with asymptotic behavior
$$
\mathcal{H}_p(v) \simeq \frac{1}{\sqrt{v}}\, e^{iv-i\frac{\pi}{4}}\ ,
\ \ \ \ \ \ \ \ \ 
|v|\gg 1\ .
$$ 

To solve for the scalar field $\chi$ it remains to fix the boundary
condition at the singularity
at $-E^2X^+X^- = x^2 = 1$. 
At the end of this section we will embed this three--dimensional model $\mathbb{M}^3/\Gamma$
in M--theory, and the compactification on $\mathbb{M}^3/\Gamma\times\mathbb{E}^8$ will yield the
$d=1$ cosmology of the previous sections. With this in mind we can fix the behavior of the field
$\chi$ at the singularity. From the orientifold construction there
are two possible boundary conditions. Either the field $\chi$ vanishes
at the singularity, or its normal derivative $\partial_x \chi$
does. This depends on which Type II particle the field $\chi$
describes. For simplicity of exposition we will explicitly consider
only the Dirichlet boundary condition. The other possibility is
similar.

It is convenient to expand the quantum field $\chi$ starting from
region $II$ and to write
$$
\begin{array}{rcl}
\chi &=&
e^{ips+i\frac{\pi}{4}} 
\left[ i\,A\, \mathcal{H}_p(i\omega |x|) 
+ B\, \mathcal{H}_p(-i\omega |x|)\right] a(p)
\spa{0.3}\\
&&+\,e^{-ips-i\frac{\pi}{4}} 
\left[ i\,A\, \mathcal{H}_p(i\omega |x|) 
+ B\, \mathcal{H}_p(-i\omega |x|)\right] a^\dagger(p)\ ,
\end{array}
$$
where $a(p)$ and $a^\dagger(p)$ are annihilation and creation
operators. The constants $A$ and $B$ are chosen real and positive and are normalized so that
$$
A^2-B^2 = 1\ .
$$
Moreover, the Dirichlet boundary condition gives
$$
\frac{B}{A} = \left|\frac{ \mathcal{H}_p(i\omega)}{\mathcal{H}_p(-i\omega)}\right|\ .
$$
We need to continue the quantum field above to the contracting and
expanding regions $I$. This involves an analytic continuation of $i|x|$
to $\pm t$. In particular, we will choose the continuation $i|x|\to
t+i\epsilon$ for the function multiplying the annihilation operator
$a(p)$ and the opposite continuation $-i|x|\to t-i\epsilon$ for the
function associated with $a^\dagger(p)$ \cite{Hawking}. 
Note that we must introduce an
infinitesimal parameter $\epsilon>0$ to correctly define the function
near the branch cut of $\mathcal{H}_p(v)$ on the negative real axis.
With such prescription the field in region $I$ has the form
$$
\begin{array}{rcl}
\chi &=& 
e^{ipz+i\frac{\pi}{4}} \left[ i\,A\, \mathcal{H}_p(\omega t+i\epsilon) 
+ B\,\mathcal{H}_p(-\omega t-i\epsilon)\right] a(p)
\spa{0.3}\\
&&+\,e^{-ipz-i\frac{\pi}{4}} \left[ i\,A\, \mathcal{H}_p(-\omega t+i\epsilon) 
+ B\, \mathcal{H}_p(\omega t-i\epsilon)  \right] a^\dagger(p)\ .
\end{array}
$$
The above choice of creation and annihilation operators defines the
quantum field starting from the singularity, and defines what we
call the intermediate vacuum, i.e. the natural vacuum for a static observer in
region $II$. 

We will now show that the constants $A$ and $B$ are the Bogolubov
coefficients which relate the intermediate vacuum to the natural
vacua of the contracting and expanding phases in region $I$. 
Firstly consider the contracting region in the far past $t\ll -1$,
where the field $\chi$ becomes
\begin{equation}
\chi \simeq \frac{1}{\sqrt{|\omega t|}}  
\left[ e^{i(\omega t+pz)}\,a_{\,\rm in}(p) 
+ e^{-i(\omega t+pz)}\, a_{\,\rm in}^\dagger(p) \right]\ ,
\end{equation}
with
\begin{equation*}
\begin{array}{c}
\displaystyle{a_{\,\rm in}(p) = 
A\,a(p) + B\, a^\dagger(-p)}\ ,
\spa{0.2}\\
\displaystyle{a^\dagger_{\,\rm in}(p) =
B\, a(-p)+A\,a^\dagger(p) }\ .
\end{array}
\end{equation*}
Similarly, in the expanding region for $t\gg 1$ the field is given by
\begin{equation}
\chi \simeq \frac{1}{\sqrt{\omega t}}  
\left[ i\,e^{i(\omega t+pz)}\,a_{\,\rm out}(p) -i\, 
e^{-i(\omega t+pz)}\, a_{\,\rm out}^\dagger(p) \right]\ ,
\end{equation}
where the creation and annihilation operators read
\begin{equation*}
\begin{array}{c}
\displaystyle{a_{\,\rm out}(p) = 
A\,a(p) - B\, a^\dagger(-p)}\ ,
\spa{0.2}\\
\displaystyle{a^\dagger_{\,\rm out}(p) = - B\, a(-p)+A\,a^\dagger(p)}\ .
\end{array}
\end{equation*}
Therefore we conclude that 
\begin{equation*}
\begin{array}{c}
\displaystyle{a_{\,\rm out}(p) = 
\alpha\,a_{\,\rm in}(p) - \beta\, a^\dagger_{\,\rm in}(-p)}\ ,
\spa{0.2}\\
\displaystyle{a^\dagger_{\,\rm out}(p) =-\beta\, a_{\,\rm in}(-p) +
\alpha\,a^\dagger_{\,\rm in}(p)}\ ,
\end{array}
\end{equation*}
where
$$
\alpha = A^2 + B^2\ ,\ \ \ \ \ \ \ \ \ \ \ \beta = 2AB\ .
$$
The natural choice for the cosmological vacuum is the past vacuum
$\left|0\right>$ defined by $a_{\,\rm in}(p)\left|0\right>=0$. Hence,
the observer in the expanding universe will detect an average number
$N(p)$ of particles of momentum $p$ given by
$$
N(p) = |\beta|^2\ .
$$
Notice that these particles arise from the reflection on the
orientifold singularity. Indeed the non--trivial coefficient $\beta$
in the Bogolubov transformation relates states with equal and opposite
momentum. Moreover, we can define an effective dimensionless temperature
$$
\frac{1}{{\mathcal T}(\omega)} = 
\frac{1}{\omega}\,\ln\left|\frac\alpha\beta\right|^2 = 
\frac{2}{\omega}\, \ln \frac{1}{2}
\left(\left|\frac{ \mathcal{H}_p(i\omega)}{\mathcal{H}_p(-i\omega)}\right|+
\left|\frac{\mathcal{H}_p(-i\omega)}{\mathcal{H}_p(i\omega)}\right|\right)\ .
$$
For $\omega\gg 1,\,m/E$ the effective temperature approaches a constant
$$
{\mathcal T} \simeq \frac{1}{2\pi}\ .
$$
In this limit ${\mathcal T}$ has the standard form
$\kappa/2\pi$, where $\kappa=1$ is either the surface gravity of the
horizon with respect to the killing vector field $\partial_s$ or the
acceleration of the orientifold mirror in the dimensionless
coordinates $EX^{\pm}$. 

\begin{figure}
\centering
\includegraphics[width=6cm]{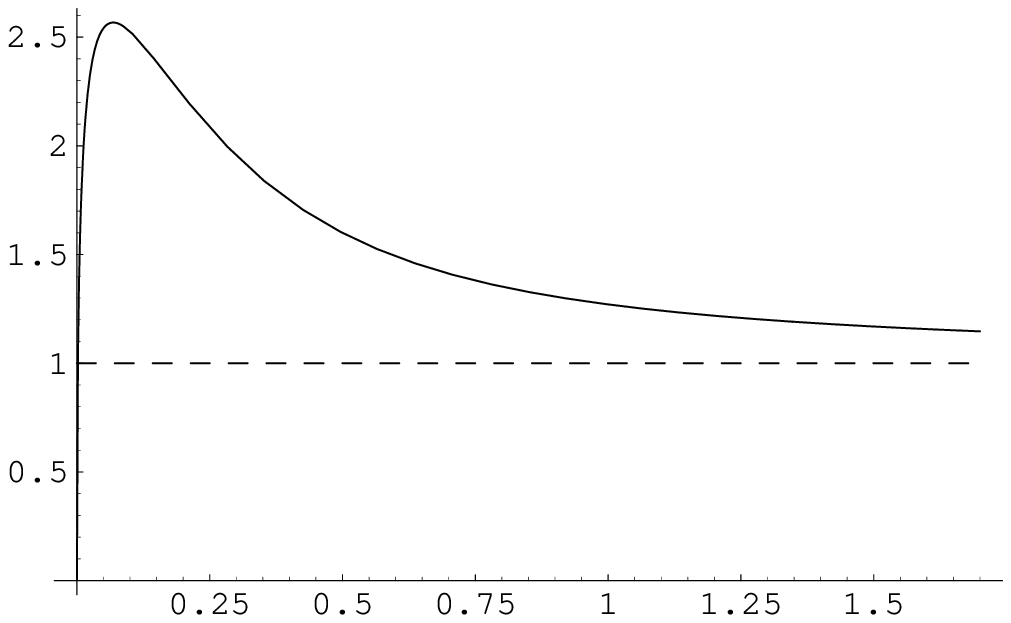}
\put(-178,115){$2\pi{\mathcal T}$}
\put(5,3){$\omega$}
\caption{\small{The effective temperature $2\pi{\mathcal T}$ as a function of
$\omega$.}}
\end{figure}

As we already mentioned, to embed this construction in the $d=1$
string theory cosmological model of the previous sections we need only
to add eight spectator flat directions and consider the M--theory
compactification on $({\mathbb M}^3/\Gamma)\times {\mathbb E}^8$. We
concentrate on the massless M--theory fields and therefore we set
$m=0$. Then the Einstein metric and the other fields are
\begin{equation*}
\begin{array}{c}
\displaystyle{
E^2ds^2=\Lambda^{\frac{1}{8}}\left[-dt^2+ds^2(\mathbb{E}^8)\right]
+\Lambda^{-\frac{7}{8}}t^2dz^2}\ ,
\spa{0.3}\\
\displaystyle{e^{\phi}=g_s\,\Lambda^{\frac{3}{4}}
=g_s\left(1+t^2\right)^{\frac{3}{4}}\ ,
\ \ \ \ \ \ \ \ \ \ 
F=-\frac{2}{g_sE}\,d\Lambda^{-1}\wedge\,dz}\ .
\end{array}
\end{equation*}
The eleven--dimensional Klein--Gordon
equation for the massless scalar field $\Xi$ reduces to 
$$
\mysquare\chi=0\ ,
$$
which is equation (\ref{KG}) in the
lorentzian polar coordinates (\ref{polar}) for $m=0$.

The comoving cosmological observer will measure a red--shifted local
energy given by  
$$
\Omega=\frac{\omega}{\sqrt{-g_{tt}}}=
\Lambda^{-\frac{1}{16}}\,E\omega\ ,
$$
and a physical temperature
$$
T=\Lambda^{-\frac{1}{16}}E\,{\mathcal T}(\omega)\ .
$$
For fixed $\Omega$ at large cosmological times, the associated
dimensionless frequency $\omega$ is blue shifted and we obtain an
exact thermal spectrum with temperature
\begin{equation*}
T=
\frac{E}{2\pi\,a(\tau)}\ ,
\end{equation*}
where $a(\tau)$ is the scale factor and $\tau$ the proper cosmological
time. 

We can extend this result to the $(d+1)$--dimensional
cosmological solutions of the previous sections. To do this 
one just needs to compute the surface gravity of the cosmological horizon
with respect to the appropriate Killing vector field. The natural
Killing vector fields to use are the generators of the $SO(1,d)$
isometries. One has to be careful with the normalization of the vector
field which is space--like in region $I$. With this in mind, one
arrives at the general result
\begin{equation}
T=\frac{{\mathcal E}}{2\pi\,a(\tau)}\ .
\label{temp}
\end{equation}
This is expected for radiation in a
$(d+1)$--dimensional cosmology, since from Boltzmann law $\rho\sim T^{d+1}$ we
obtain $\rho\sim 1/a^{d+1}(\tau)$, which follows from the radiation
equation of state $\rho=d\,p$.

\section{Conclusion}

In this work we show that the presence of a localized negative cosmological constant
on a time--like lower dimensional hypersurface avoids the conventional
space--like singularity of the big--crunch/big--bang scenarios. This
localized energy density arises naturally in string theory as branes with
negative tension like orientifolds. The resulting non--trivial cosmological
background has a horizon, which is created naturally in orientifold models
with broken supersymmetry. Space--time is divided in three regions. Firstly,
there is a collapsing region with a future cosmological horizon. As we pass
this horizon, we enter a `static' region with the localized
orientifolds. This region has a future horizon. Crossing this horizon we
enter the expanding phase. The key idea is that the existence of the horizon
flips the would be space--like singularity to a time--like singularity
leading to a brane resolution of the singularity. 

To study the generic properties of the proposed cosmological scenario we
consider a specific model in string theory. The interpretation of the
singularity as an orientifold was justified by analyzing the geometry
near the singularity and by probing it 
with D--branes. The analysis of this effective potential showed that there
is an energy density at the core of the geometry, which prevents the
orientifolds to collapse due to the gravitational repulsion. Also, the
resolution of the singularity provides well defined boundary conditions,
which are necessary to determine past/future transition amplitudes. In
particular, we considered the vacuum to vacuum amplitude. Assuming the
trivial vacuum in the past, we derived a thermal spectrum for
radiation in the far future with temperature given by equation
({\ref{temp}).

\section*{Acknowledgments}
We would like to thank Carlo Angelantonj, Carlos Herdeiro, Elias Kiritsis
and Rodolfo Russo for fruitful discussions. 
L.C.  would like to thank the Ecole Normale
Sup\'erieure and M.S.C. the Instituto
Superior T\'ecnico in Lisbon for hospitality. 
This work was partially supported by European Union under the
RTN contracts HPRN--CT--2000--00122 and --00131. The work
of M.S.C. is supported by a Marie Curie Fellowship under the European
Commission's Improving Human Potential programme (HPMFCT--2000--00508).

\end{document}